# Using an Impact Hammer to Perform Biomechanical Measurements during Osteotomies: Study of an Animal Model

Léo Lamassoure[a], Justine Giunta[b], Giuseppe Rosi[c], Anne-Sophie Poudrel[a], Romain Bosc[b,d], and Guillaume Haïat[a].

[a] *CNRS, Laboratoire Modélisation et Simulation Multi Echelle, MSME UMR 8208 CNRS, 61 avenue du Général de Gaulle, 94010 Créteil Cedex, France.*

[b] *Service de Chirurgie Plastique, reconstructrice, esthétique et Maxillo-faciale du Centre Hospitalier Universitaire Henri Mondor, 51 avenue du Maréchal de Lattre de Tassigny, 94000 Créteil, France ;*

[c] *Université Paris-Est, Laboratoire Modélisation et Simulation Multi Echelle, MSME UMR 8208 CNRS, 61 avenue du Général de Gaulle, 94010 Créteil Cedex, France.*

[d] *Équipe 10, Groupe 5, IMRB U955, INSERM/UPEC, 51 Avenue du Maréchal de Lattre de Tassigny, 94010 Créteil, France*

[*]**Corresponding author:**
Guillaume HAÏAT
Laboratoire Modélisation et Simulation Multi Echelle, MSME UMR CNRS 8208,
61 avenue du Général de Gaulle,
94010 Créteil, France
Tel: (33) 1 45 17 14 41
Fax: (33) 1 45 17 14 33
E-mail: **guillaume.haiat@cnrs.fr**

*Abstract*

Osteotomies are common surgical procedures used for instance in rhinoplasty and usually performed using an osteotome impacted by a mallet. Visual control being difficult, osteotomies are often based on the surgeon proprioception to determine the number and energy of each impact. The aim of this study is to determine whether a hammer instrumented with a piezoelectric force sensor can be used to i) follow the displacement of the osteotome and ii) determine when the tip of the osteotome arrives in frontal bone, which corresponds to the end of the osteotomy pathway.

Seven New Zealand White rabbit heads were collected, and two osteotomies were performed on their left and right nasal bones using the instrumented hammer to record the variation of the force as a function of time during each impact. The second peak time $\tau$ was derived from each signal while the displacement of the osteotome tip $D$ was determined using video motion tracking.

The results showed a significant correlation between $\tau$ and $D$ ($\rho^2 = 0{,}74$), allowing to estimate the displacement of the osteotome through the measurement of $\tau$. The values of $\tau$ measured in the frontal bone were significantly lower than in the nasal bone ($p < 10^{-10}$), which allows to determine the transition between the nasal and frontal boneswhen $\tau$ becomes lower than 0.78 its initial averaged value.

Although results should be validated clinically, this technology could be used by surgeons in the future as a decision support system to help assessing the osteotome environment.

## I. INTRODUCTION

Osteotomes are used in various disciplines such as orthopedic, maxillofacial and plastic surgery. They are employed in combination with a hammer, or mallet, following the same principle as a chisel, and their function is essentially to cut bone and cartilage tissues. For instance, in maxillo-facial surgery, they are used to cut the maxilla bone or split and expend the bone ridge to improve bone density and primary stability of dental implants (1,2). In orthopedic surgery, osteotomies are performed to cut bone tissue and to shave off osteoperiosteal grafts as well as to remove the cartilage and subchondral bone (3). Osteotomes are also important tools in plastic surgery (4) and in particular for nose surgery where it is used to correct nasal deviation, narrow the bony nasal vault, or reduce the dorsal hump (5,6). Despite their frequent clinical use, osteotomies are still associated with a certain number of risks. An important issue lies in that the surgeon usually does not have any visual feedback on the state of the targeted tissue during the osteotomy. Various technologies allowing image-guided surgery (e.g. using PET, SPECT, DTI, fMRI, or electrophysiology) have been developed and are now used in the operating room (7,8). However, they are usually not adapted to osteotomies because of their complexity or radiating features. Moreover, the aforementioned techniques imply an important increase of the duration of the surgical procedure. Therefore, most surgeons still rely so far on their proprioception and experience to perform osteotomies.

Biomechanical approaches represent interesting modalities because they can be used to retrieve information about the properties of the tissue surrounding the osteotome tip, which could be used by the surgeon as a decision support system during the procedure. For example, in rhinoplasty, nasal bone tissues are thicker where they connect to frontal bones and thinner in their inferior part where they articulate with the maxilla and the upper lateral cartilage (9). An estimation of bone thickness during the osteotomy could therefore allow the surgeon to locate more precisely the tip of the osteotome. Such estimation is important during rhinoplasty since errors on the osteotome pathway could result in permanent deformities (9,10). Moreover, an estimation on the progress of the osteotome position would also be of interest, as well as any information about the mechanical properties of the tissue in contact with the osteotome, which could be used to help the surgeon adjust the impact energy and to detect the apparition and propagation of cracks. Incomplete fractures are sometimes preferable in older patients to avoid uncontrolled cracks, which can cause a collapse of the nasal vault, leading to an inverted V deformity (5,9,11). All these phenomena may be particularly difficult to detect, especially for young surgeons (12), and additional information on the properties of the osteotomy site could provide an interesting insight and potentially reduce the failure risks of these procedures.

Over the last few years, our group has developed a technology based on impact analyses aiming at measuring implants stability. It consists in a hammer instrumented with a piezoelectric force sensor, which measures the variation of the force as a function of time during each impact. The results showed that this approach can be used to assess the primary stability of acetabular cup implants and the insertion endpoint of femoral stems. Various indicators derived from the signal (13–15) were tested *in vitro*, and then *ex vivo* (16-18). Static (16,17) and dynamic (18) finite element models were developed to better understand the mechanical phenomena occurring during implant insertion. Eventually, pre-clinical studies were carried out on anatomical subjects to validate the device under more realistic conditions (22,23).

Based on the results obtained in the context of hip implants, the same approach consisting in using an instrumented hammer was considered in a preliminary *in vitro* study aiming at investigating the possible applications of this technique to osteotomies (19). The objective was to determine whether information about the properties of a sample impacted by the osteotome could be retrieved using an instrumented hammer. One hundred samples made of different composite materials and with various thicknesses were considered and a model-based inversion technique was developed based on the analysis of two indicators derived from the analysis of the variation of the force as a function of time. This study, which also included an analytical model, showed that it was possible to i) classify the samples depending on their material types, ii) determine the materials stiffness and iii) estimate the samples thicknesses, which opened the path for the application of this approach to preclinical testing using an animal model.

The aim of the present study is to determine whether an instrumented hammer can be employed in an animal model in order to i) determine the displacement of the osteotome during an impact and ii) estimate when the tip of the osteotome enters frontal bone tissue. To do so, our approach is to use the rabbit heads, which is a validated animal model for rhinoplasty (20).

## II. Materials and Methods

### A. Sample Preparation

The New Zealand White rabbit was chosen for the animal model since it is commonly used to model osteotomies (21,22) and because it presents similarities with the human anatomy (20). Seven rabbit heads were obtained from the National veterinary school of Alfort (Maisons-Alfort, France). They were collected in the 24 hours following the animal's euthanasia and were then frozen for storage. They were taken out of the freezer one day before the experiment and were thawed at room temperature to allow them to be completely unfrozen by the time the experiment was performed. The study has been approved by the ethical committee of the Alfort National Veterinary School (ENVA). The sample preparation followed several steps. First, the surrounding soft tissues were carefully scrapped away to reveal parts of the supraoccipital and squamosal bones in order to be able to hold the sample properly. Second, the posterior part of the head was embedded in a fast hardening resin (Smoothcast 300, Smooth-On, Easton, PA, USA) in order to be able to hold the sample in a vise, following the procedure followed in (23). Third, the snout was excised, and an incision was performed along the nose's axis. The skin flaps were pulled aside in order to reveal the nasal bone, the anterior part of the frontal bone, and the upper part of the premaxilla, as shown in Fig. 1.

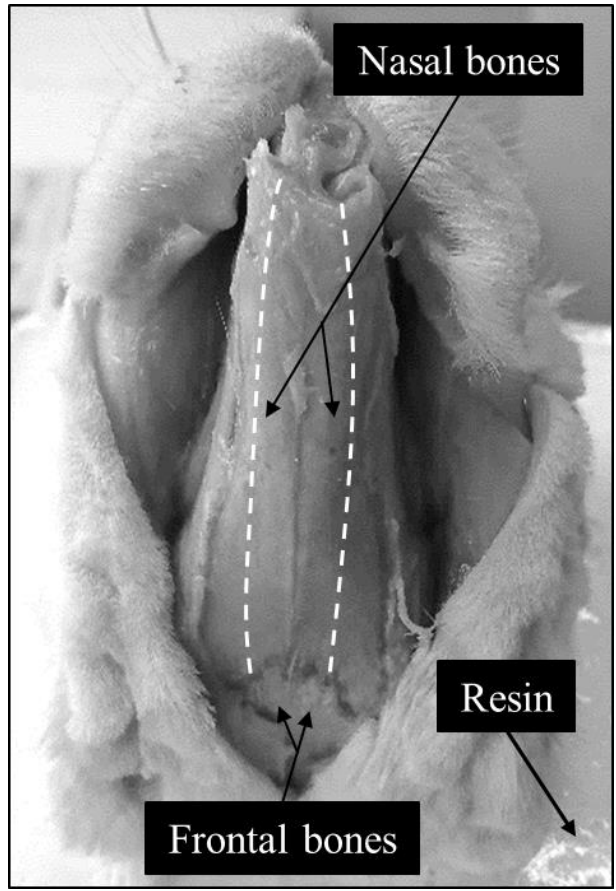

**Fig. 1.** Image of the sample corresponding to the New Zealand White Rabbit head after preparation and just before the beginning of the experiment. The head was embedded into the resin and held in a vise. The nasal bones are completely uncovered, and the planned osteotomy pathways are represented by white dashed lines.

*B. Instrumented Hammer*

A 260 g surgical mallet (32-6906-26, Zepf, Tuttlingen, Germany) was used in this study to impact an osteotome (32-6002-10, Zepf, Tuttlingen, Germany) with a 10 mm long cutting edge. The hammer and the osteotome are the same as the ones used in clinics. For each impact, the osteotome was held manually, similarly to what is done in the clinic. A dynamic piezoelectric force sensor (208C04, PCB Piezotronics, Depew, NY, USA) with a measurement range up to 4.45 kN in compression was screwed in the center of the impacting face of the hammer to measure the force applied to the osteotome. A data acquisition module (NI 9234, National Instruments, Austin, TX, USA) with a sampling frequency of 51.2 kHz and a resolution of 24 bits was used to record the time variation of the force exerted on the osteotome. The data were transferred to a computer and recorded with a LabVIEW interface (National Instruments, Austin, TX, USA) for a duration of 2 ms. A schematic description of the experimental setup is presented in Fig. 2.

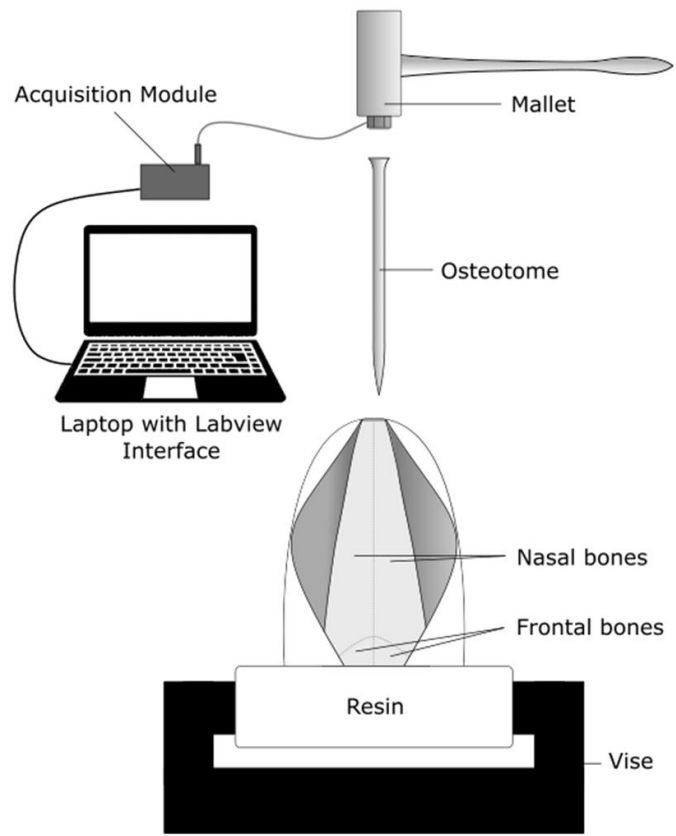

Fig. 2. Schematic description of the experimental setup.

*C. Experimental Procedure*

For each rabbit, an osteotomy was performed on each nasal bone (right and left), resulting in a total of 14 osteotomy pathways. Both osteotomy pathways are indicated in Fig. 1. The experiments were realized by an experienced maxillofacial surgeon. The position of the osteotome tip was filmed using a camera (L-920M3, Spedal, Taiwan) during the entire experiment.

The osteotomies were set to begin at the anterior end of the nasal bone and consisted in consecutive hammer impacts of the instrumented hammer on the osteotome having its blade in contact with bone tissue. The variation of the force as function of time was recorded by the force sensor for each impact and the maximum value of the force was noted $F$. The following protocol was followed by the surgeon.

A series of approximately 10 impacts was realized with a relatively low energy, the value of $F$ for each impact being comprised between 100 and 400 N. The corresponding impacts were denoted as "weak" in what follows. Then, a series of around 10 impacts was realized with a relatively high energy, the value of $F$ for each impact being comprised between 800 and 2000 N. The corresponding impacts were denoted as "strong" in what follows. Then, the aforementioned procedure was repeated so that series of "weak" impacts were followed by series of "strong" impacts and conversely. The series of strong impacts were used to fracture bone tissue and induce a displacement of the osteotome. Crack propagation may occur, leading to a movement of the osteotome in bone tissue during the series of strong impacts. The aim of the series of strong impact was to determine whether the instrumented hammer could be used to determine the displacement of the osteotome due to crack initiation. In turn, the aim of realizing a series of weak impacts was to determine whether such approach, which has the advantage of being noninvasive since it is not likely to modify the position of the osteotome, could be used to determine whether the osteotome was located in nasal or frontal bone. The choice of the range of variation chosen for F for the two series of impact will be discussed in section 4. Note that the surgeon had only access to the value of $F$ during the experiments to help them stay within the selected range, but not to the value of $\tau$ and $D,$ that was obtained by post-processing.

The alternation of the strong and weak impact series was pursued until the tip of the osteotome appeared to have reached the frontal bone, which was determined empirically by the surgeon using his proprioception. In particular, similarly as what is done in the clinic, the surgeon relied on his sight (an open procedure is used and bone tissue is fully visible) and his hearing, since the arrival of the osteotome on the frontal bone is associated with a subtle change of the noise produced during the impact.. In order to study the transition between nasal and frontal bone, a dozen additional impacts were performed to obtain measurements in frontal bone before ending the trial, so that the last series of impact corresponded to weak impacts.

### D.  Data Processing

Data processing was done following the same method as the one used for the previous study (19). The time dependence of the force applied to the osteotome during a given impact was measured with the force sensor described above, leading to a signal denoted $s(t)$. Two examples of typical signals are shown in Fig. 3. A dedicated signal processing technique was applied to $s(t)$ using information derived from the different peaks obtained in the signal. For the first ($p$=1) and second ($p$=2) peaks of $s(t)$, the maximum peak amplitude $a_p$, the time of its center $t_p$, and the root mean square width $w_p$ were determined by fitting a Gaussian function with a time windows centered on the interval corresponding to $s(t) > a_p/5$. The second peak time $\tau$ was defined based on the difference between the times of the second and first peaks of $s(t)$. Figure 3 shows the resulting value of $\tau$ for two typical signals.

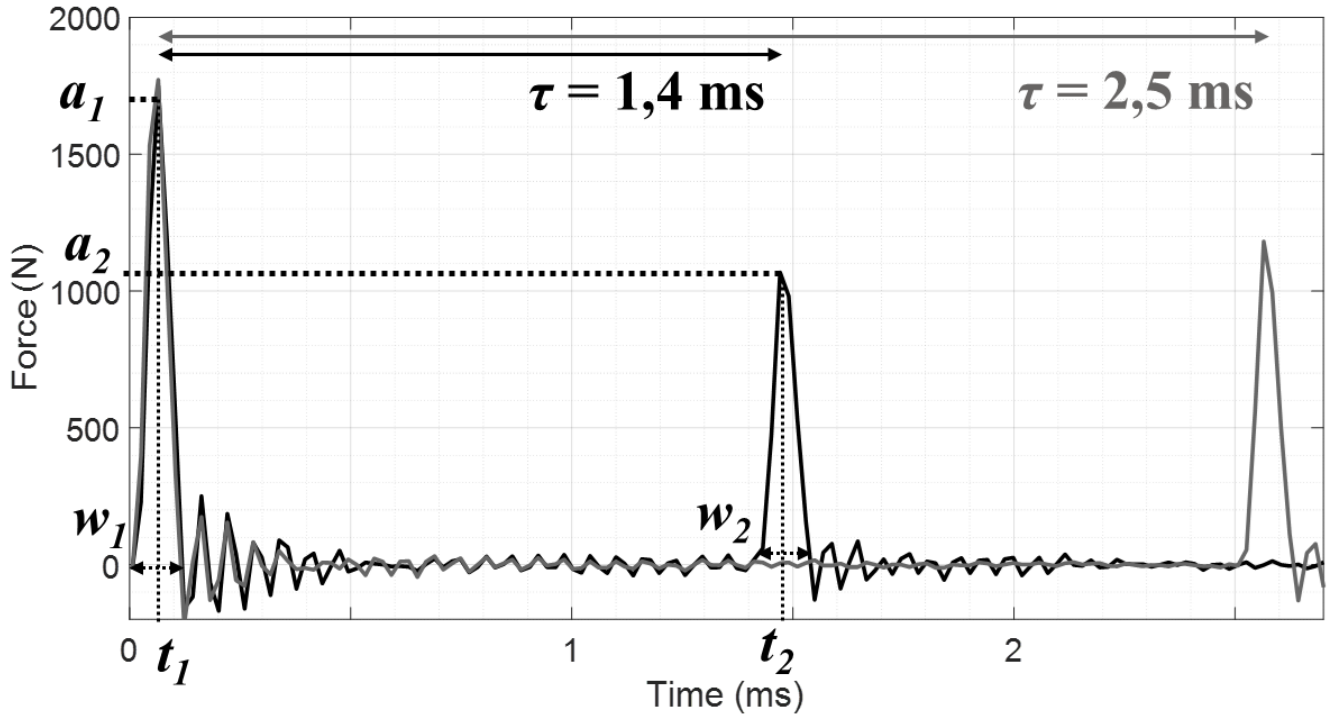

**Fig. 3.** Example of two typical signals *s(t)* corresponding to the variation of the force as a function of time obtained during an impact (strong series) of the instrumented hammer on the osteotome. The corresponding values of the parameter $\tau$ are given. The different parameters of the signal are indicated for the signal in black: $a_1$ and $a_2$ are the amplitudes of the first and second peak, respectively; $w_1$ and $w_2$ are the root mean square width of the first and second peak, respectively; $t_1$ and $t_2$ are the time of the center of the first and second peak, respectively.

The average and standard deviation of all values of $\tau$ was determined for each osteotome pathway when the osteotome tip was in nasal (respectively frontal) bone and were noted $\tau_n^m$ and $\tau_n^{SD}$ (respectively $\tau_f^m$ and $\tau_f^{SD}$). For each osteotomy pathway, $M_\tau$ was defined as the averaged value of $\tau_n^m$ obtained for the first two series of weak impacts for the corresponding osteotomy pathway.

Based on the video of the sample and of the osteotome, a parameter $D$ was defined as the absolute displacement of the tip of the osteotome corresponding to a given strong impact. To do so, a tracker software (Tracker, National Science Foundation, Alexandria, USA) was used to compare the images corresponding to the position of the osteotome relatively to the sample obtained before and after the impact (Figure 4).

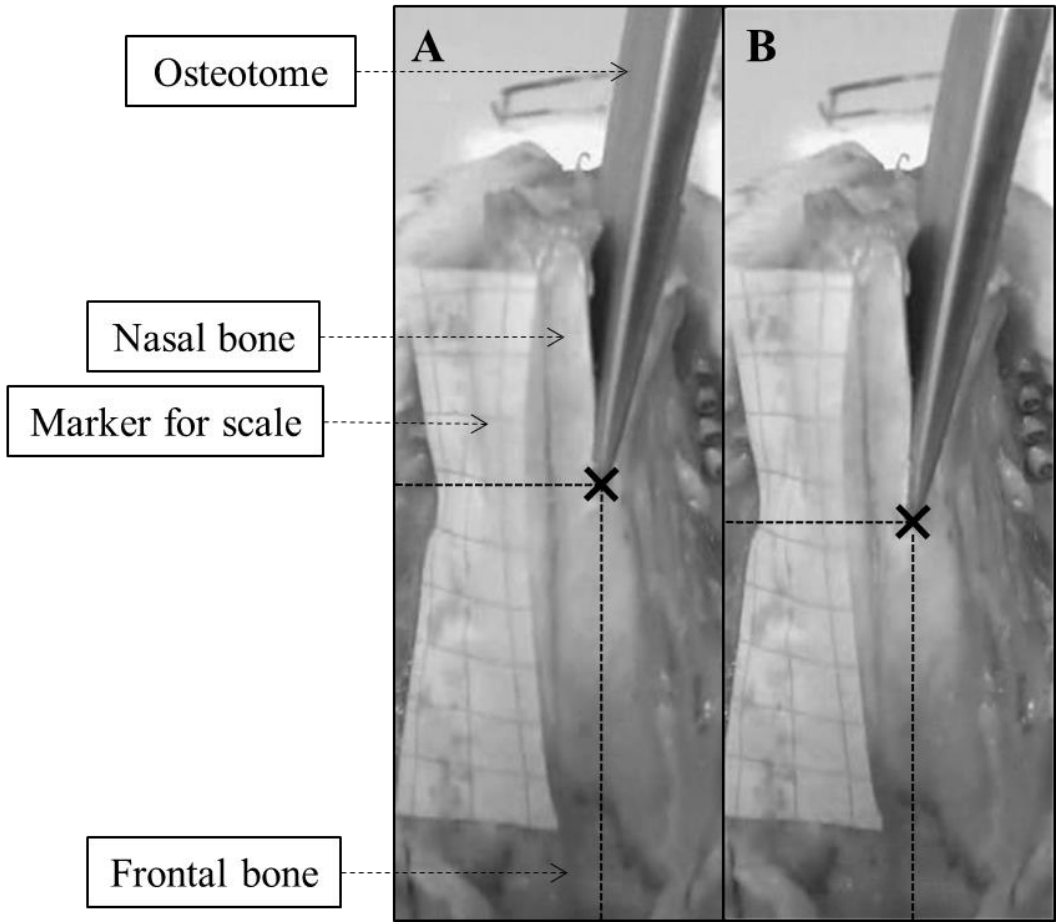

**Fig. 4.** Illustration of the measurement of displacement of the osteotome D. A: Configuration before a given "strong" impact. B: Configuration after the same impact. The position of the tip of the osteotome is indicated by a cross.

Two analyses were carried out. First, the relationship between $\tau$ and $D$ obtained for strong impacts only was analyzed using a linear regression analysis. Second, the variation of $\tau$ was analyzed according to the type of bone where the osteotome tip was located (nasal or frontal bone).

## III. RESULTS

Figure 5 shows an example of the values of $F$ and of $\tau$ obtained for a given osteotomy pathway. Figure 5A illustrates the alternating series of weak and strong impacts. The weak impacts correspond to lower and more stable values of $\tau$. The values of $\tau$ for the strong impacts are usually higher but show a sudden decrease at the arrival in the frontal bone, as illustrated in Fig. 5B for impacts #97-100. Over the full database, the average values of $\tau_n^{SD}$ (nasal bone) and of $\tau_f^{SD}$ (frontal bone) were equal to 0,2 ms and 0,3 ms, respectively, which corresponds to the reproducibility of the measurement of $\tau$ for weak impacts.

The values of $\tau$ and $D$ are compared for all strong impacts where $D$ was not equal to zero in Fig. 6. The values of $\tau$ and $D$ are found to be significantly correlated, with a Spearman correlation coefficient $\varrho^2 = 0{,}74$ ($p < 10^{-10}$).

The measurements realized during the weak impacts allowed to study the evolution of the values of $\tau$ along the osteotome pathway. Figure 7 shows the distribution of the values of $\tau$ for each osteotomy pathways when the osteotome is located in nasal and frontal bone. The values of $\tau_n^m$, $\tau_n^{SD}$, $\tau_f^m$ and $\tau_f^{SD}$ are indicated in Fig. 7. An ANOVA analysis shows that the values of $\tau$ were significantly lower in frontal bone compared to the nasal bone for all osteotomy pathway ($p<10^{-10}$). A comparison of these values can be seen on Fig. 7. For all osteotomy pathways, the osteotome tip was found to enter frontal bone when $\tau_n^m < 0{,}78M_\tau$, which therefore corresponds to a critical value below which the osteotome is located in frontal bone.

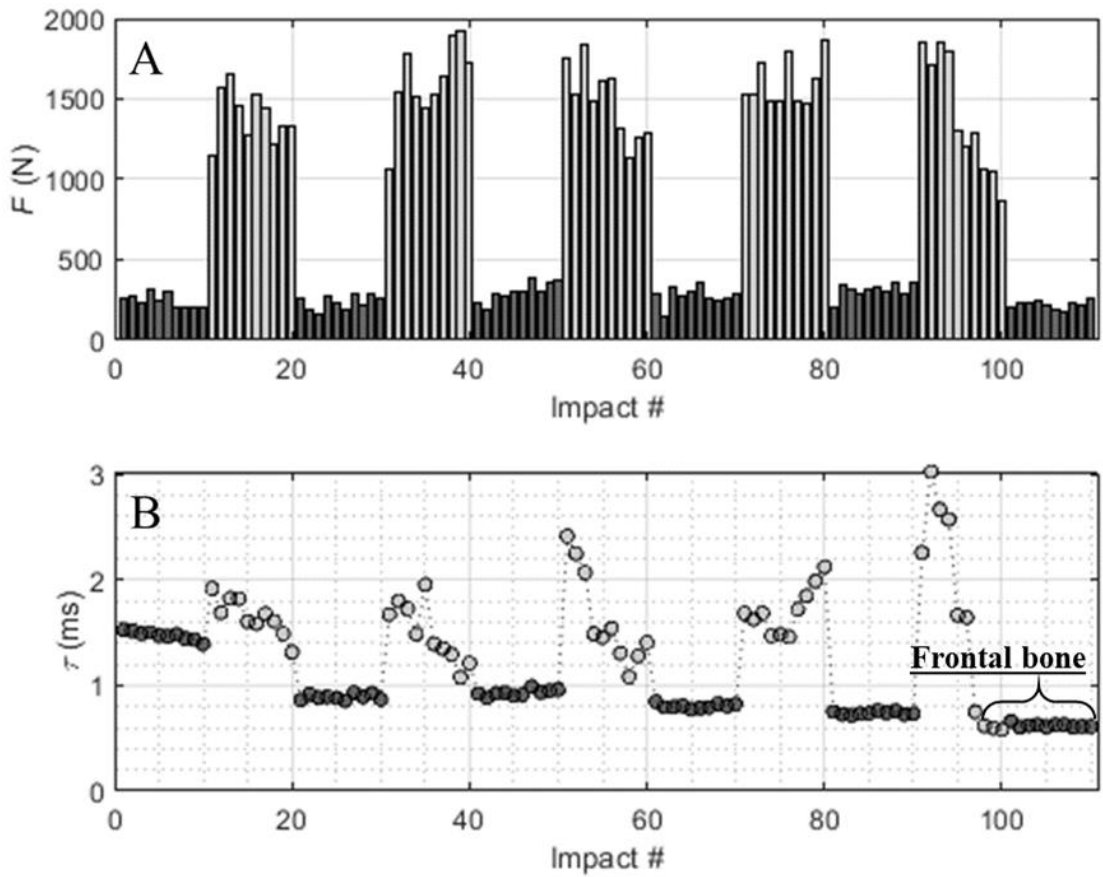

**Fig. 5.** Evolution of the maximum peak force F (A) and the parameter τ (B) for a given entire osteotomy pathway. The impacts performed after the arrival in frontal bone are indicated. For both A and B, the data is represented in dark grey for the weak impacts, and light grey for the strong impacts.

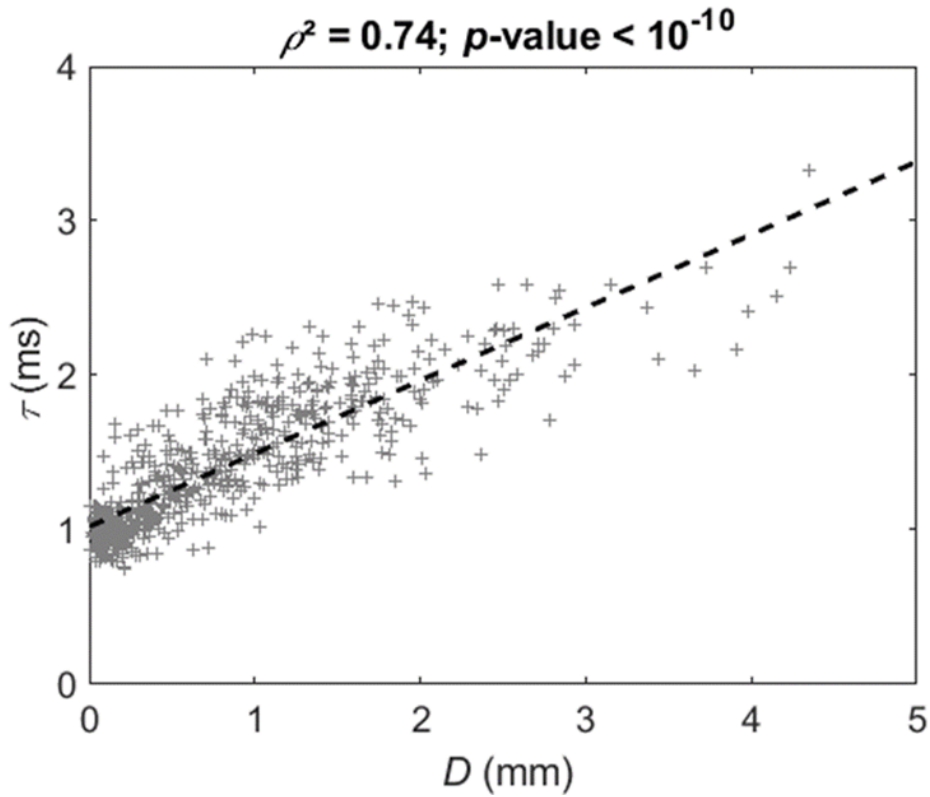

**Fig. 6.** Relationship between the parameter τ and the displacement D. The Spearman correlation coefficient $\varrho^2$ and the corresponding p-value are indicated. The dashed line corresponds to the linear regression analysis and has the following equation: $\tau = 0.4725 \times D + 1.0214$.

## IV. DISCUSSION

The originality of the present *ex vivo* study is to show that the instrumented hammer could be used to i) detect the moment when the osteotome tip arrives in frontal bone and ii) estimate the displacement of the osteotome in bone tissue due to impacts. These results are in agreement with the preliminary study realized with composite materials, which has shown that the indicator τ was sensitive to the rigidity of the material in contact with the tip of the osteotome (19).

The physical interpretation of the mechanical phenomena occurring during each impact of the hammer on the osteotome has been discussed in detail in (19), in particular thanks to the development of an analytical model. Note that finite element numerical simulations were also realized for an instrumented hammer used to assess acetabular implant stability (18). Briefly, the force applied by the hammer to the osteotome during the first contact induces an acceleration of the osteotome, which then bounces back and forth between bone tissue and the head of the hammer (which has a significantly higher mass than the osteotome). Such phenomenon leads to the different peaks of the force shown in Fig. 3 because several iterations of contact and rupture of contact occur during a single impact between the hammer and the osteotome. Therefore, the time between the first and second peak (denoted τ herein) is related to the rigidity of the material around the osteotome tip. In particular, the frontal bones of rabbit (as well as of humans) are denser and thicker than the nasal bones (24,25), which explains the higher values of $\tau$ obtained in nasal bone compared to frontal bone (see Fig. 7).

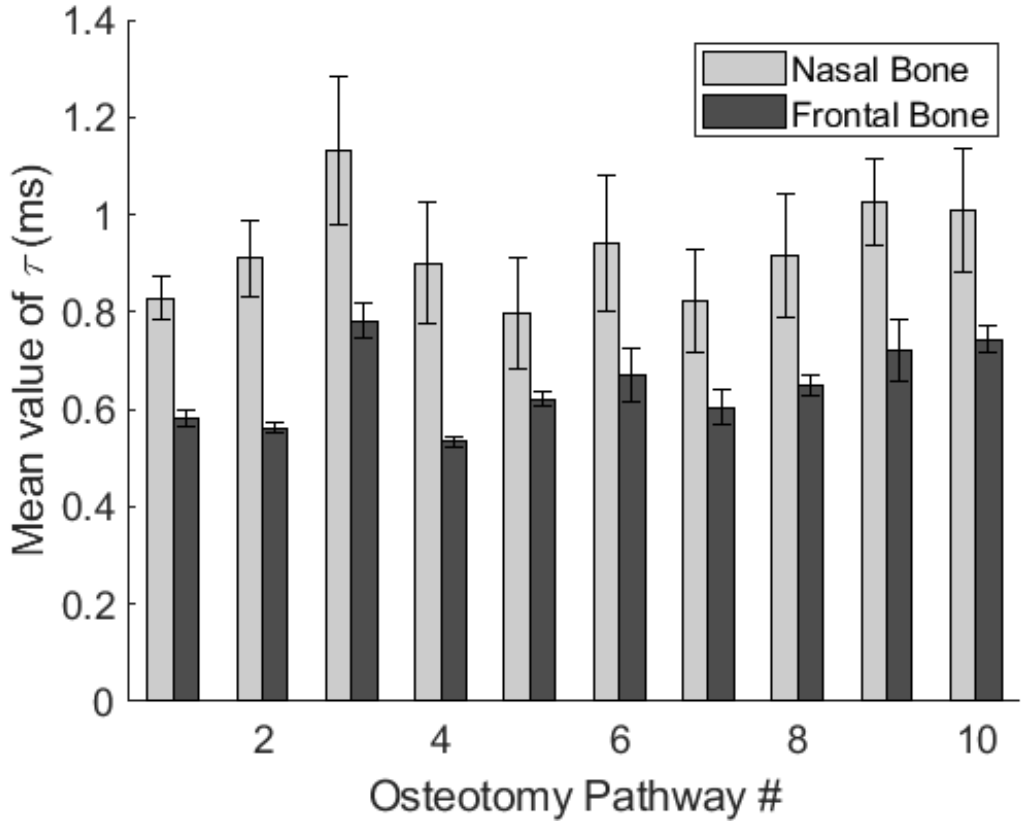

**Fig. 7.** Average and standard deviation of all values of $\tau$ for each osteotomy pathway when the osteotome tip was in nasal (respectively frontal) bone, noted $\tau_n^m$ and $\tau_n^{SD}$ (respectively $\tau_f^m$ and $\tau_f^{SD}$). The error bars correspond to standard deviation values.

The aforementioned phenomenological explanation may also explain qualitatively the correlation obtained between $D$ and $\tau$ (see. Fig. 6) because the displacement of the osteotome $D$ is related to a crack initiation due to a strong enough impact. First, crack initiation leads to a decrease of the rigidity of bone tissue located in front of the osteotome, which leads to an increase of $\tau$, as shown in (19). Second, after crack propagation in bone tissue, the osteotome shall move forward into the crack and then bounce back on the crack tip. The presence of the crack will therefore lead to a delay before the osteotome comes back to the hammer because of the time necessary to reach the crack tip. Therefore, when $D$ increases, the delay before the osteotome comes back to the hammer also increases, which leads to an increase of $\tau$. Note that this phenomenological interpretation may also explain the results shown in Fig. 5 and in particular i) the higher values of $\tau$ obtained for weak impacts (when no crack may propagate) compared to strong impacts and ii) the better reproducibility value of $\tau$ obtained for weak impacts (when no crack may propagate) compared to strong impacts. The two aforementioned phenomena concur in the positive correlation of $D$ and $\tau$ shown in Fig. 6.

The range of variation of the force obtained for weak impacts (100-400 N) was chosen to be as low as possible while being high enough to trigger the recording of an acceptable signal. The range of variation of the force obtained for strong impacts (800-2000N) was set to be similar to the one used in clinics during real rhinoplasty procedures, following the recommendation of the maxillofacial surgeons. The size of both force ranges comes from the difficulty for the operator to impact the osteotome with a precise force.

Although we could have expected that $D$ could increase as a function of the peak force (since the osteotome would be more likely to progress when the impact energy is higher), ***no correlation between*** the peak force ***and $D$*** was found (data not shown). An explanation for this absence of correlation between $D$ and the peak force is that $D$ does not only depend on the impact energy but also on the biomechanical properties of the tissue in contact with the tip of the osteotome, which seems to be the main determinant for the apparition of fractures that allow the osteotome to move forward. For each impact, the biomechanical phenomenon at work could be explained as follows. Below a certain threshold, the impact force is too low to generate a fracture, leading to a displacement of $D = 0$. Above that threshold, a fracture is created and its size is likely to depend more importantly on the properties of the tissue than the impact force. Note that $\tau$ is likely to depend both on bone mechanical properties and on bone thickness.

This study presented several limitations. First, a rabbit model has been employed, although other animal models, such as minipigs, rats, and goats, have been considered for maxillofacial surgery (26,27) and osteotomies (28,29). However, the rabbit remains the most widely animal model used for rhinoplasty (20). Note that due to ethical reasons, it was not possible to perform the experiments in anatomical subject without validating the approach in an animal model. Although fresh-frozen samples would have been the preferred solution (30) since freezing may impact bone mechanical properties (31), all samples underwent the same freezing/thawing protocol. Note that a previous study (32) using an instrumented hammer to retrieve the acetabular cup implant stability showed that the results did not depend on the amount of soft tissues located under the embedded bone sample.

Second, the 2-D view obtained with the video (see Fig. 4) induces sources of error in the determination of $D$. In order to minimize this error, the camera was always positioned with its axis normal to the surface of interest (i.e., the nasal bones). For each osteotome pathway, the error due to perspective related issues was checked using scales of know dimensions as references and was always lower than 3%.

Third, the same osteotome was used for all experiments. Surgeons actually have access to a variety of osteotomes with different geometry, which is likely to affect the results of the surgical procedure (33). The choice of the osteotome is also likely to modify the signal obtained with the instrumented hammer and further studies are required in order to assess the effect of the osteotome on the results. Moreover, the load sensor is uniaxial and only the force perpendicular to the axis of the osteotome was measured. The surgeon was instructed to impact the osteotome in the direction of its axis and an angle between the osteotome axis and the impact direction may cause discrepancy of the measurement. The influence of such an angle should be investigated in future works.

## V. Conclusion

The long-term objectives of this research are to (1) detect a change of the tissue material properties during an osteotomy, (2) determine the stiffness of the bone tissue around the osteotome tip, and (3) estimate the thickness of the materials being osteotomized. These three pieces of information may be used to assist surgeons during osteotomy procedures via a decision support system.

The results obtained with this *ex vivo* animal model suggest that the instrumented hammer could be used as a decision support system to assist surgeons during osteotomy procedures. In particular, we focused on i) the detection of the transition between nasal and frontal bone and ii) the correlation between the indicator $\tau$ obtained using the instrumented hammer and the displacement of the osteotome during the corresponding impact.

Future studies should consider anatomical subjects similarly as what was done in (34,35) in the context of hip implants, in order to confirm the results obtained for i) the correlation between $\tau$ and $D$ and ii) the possibility to detect the arrival of the osteotome tip on frontal bone. The advantage of cadaveric experiments is that it will be possible to measure $D$ and to work with an open surgery. The tool should then be calibrated using cadaveric experiments and machine learning technique can also be considered, similarly as what was done in (19). Once the tool is actually calibrated in cadaveric experiments, clinical trials will be carried out and the results will then be compared to the surgeon proprioception and, if possible, to video motion tracking. Moreover, it will also be necessary to focus on the phenomena responsible for crack initiations.

## Acknowledgment

This project has received funding from the European Research Council (ERC) under the European Union's Horizon 2020 research and innovation program (grant agreement No 682001, project ERC Consolidator Grant 2015 BoneImplant).

The authors would like to acknowledge the support of the "Prématuration programme" of the CNRS through the Osteome project.